\documentclass[sigplan,nonacm]{acmart}
\usepackage{gensymb}
\usepackage[normalem]{ulem}
\usepackage{xcolor}
\usepackage{amsmath,amsfonts}
\usepackage{algorithmic}
\usepackage{textcomp}
\usepackage{enumitem}
\usepackage{listings}
\usepackage{subcaption}
\usepackage{soul}
\usepackage{comment}
\usepackage{tikz}
\usepackage{xspace}
\usepackage{array}
\hypersetup{
    unicode=true,
    bookmarksnumbered=true,
    colorlinks=true,
    linkcolor={red!70!black},
    citecolor={red!70!black},
    urlcolor={blue!70!black},
    pdfborder={0 0 0}
}
\usepackage[capitalize,nameinlink]{cleveref}
\usepackage{flushend}
\usepackage{pifont}

\usepackage{graphicx}
\usepackage{array}

\newcommand{\myparagraph}[1]{\vspace{\smallskipamount}\noindent\textbf{#1.\xspace}}

\newcommand{\myparagraphemph}[1]{\vspace{\smallskipamount}\noindent\emph{#1.\xspace}}
\newcommand{\eg}{\emph{e.g.}\xspace}

\newcommand{\ie}{\emph{i.e.}\xspace}

\newcommand*{\rom}[1]{\uppercase\expandafter{\romannumeral #1\relax}}
\newcommand{\system}{TAPAS}
\newcommand{\provider}{\emph{Azure}}

\newcommand{\greenuparrow}{\textcolor{green}{$\boldsymbol{\uparrow}$}}
\newcommand{\reddownarrow}{\textcolor{red}{$\boldsymbol{\downarrow}$}}
\newcommand{\reduparrow}{\textcolor{red}{$\boldsymbol{\uparrow}$}}
\newcommand{\greendownarrow}{\textcolor{green}{$\boldsymbol{\downarrow}$}}

\newcounter{insightcounter}
\newcommand{\myinsight}{%
    \refstepcounter{insightcounter}
    \vspace{\smallskipamount}\noindent\emph{\underline{Insight \#\theinsightcounter:\xspace}}%
}

\newif\ifshowcomment
\showcommenttrue

\ifshowcomment
    \newcommand{\jovan}[1]{{\color{blue}[JS: #1]}}
    \newcommand{\inigo}[1]{{\color{blue}[IG: #1]}}
    \newcommand{\chaojie}[1]{{\color{magenta}[CZ: #1]}}
    
    \newcommand{\hq}[1]{{\color{brown}[HQ: #1]}}
        
    \newcommand{\todo}[1]{{\color{red}[TODO: #1]}}
\else
    \newcommand{\jovan}[1]{\ignorespaces}
    \newcommand{\inigo}[1]{\ignorespaces}
    \newcommand{\chaojie}[1]{\ignorespaces}
    \newcommand{\hq}[1]{\ignorespaces}
    \newcommand{\todo}[1]{\ignorespaces}
\fi

\title{\system: Thermal- and Power-Aware Scheduling for LLM Inference in Cloud Platforms}

\author[J. Stojkovic, C. Zhang, Í. Goiri, E. Choukse, H. Qiu, R. Fonseca, J. Torrellas, and R. Bianchini]{
    Jovan Stojkovic\textsuperscript{\textdagger}$^1$
    \quad
    Chaojie Zhang
    \quad
    Íñigo Goiri
    \quad
    Esha Choukse
    \\
    Haoran Qiu
    \quad
    Rodrigo Fonseca
    \quad
    Josep Torrellas\textsuperscript{\textdagger}
    \quad
    Ricardo Bianchini\textsuperscript{\textdaggerdbl}
}

\affiliation{
    \textit{Microsoft Azure Research}
    \qquad
    \textsuperscript{\textdaggerdbl}\textit{Microsoft Azure}
    \qquad
    \textsuperscript{\textdagger}\textit{University of Illinois at Urbana-Champaign}
    \country{USA}
    \vspace{5mm}
}

\date{}

\begin{document}

\begin{abstract}
The rising demand for generative large language models (LLMs) poses challenges for thermal and power management in cloud datacenters.
Traditional techniques often are inadequate for LLM inference due to the fine-grained, millisecond-scale execution phases, each with distinct performance, thermal, and power profiles.
Additionally, LLM inference workloads are sensitive to various configuration parameters (\eg, model parallelism, size, and quantization) that involve trade-offs between performance, temperature, power, and output quality.
Moreover, clouds often co-locate SaaS and IaaS workloads, each with different levels of visibility and flexibility.

We propose \system{}, a thermal- and power-aware framework designed for LLM inference clusters in the cloud.
\system{} enhances cooling and power oversubscription capabilities, reducing the total cost of ownership (TCO) while effectively handling emergencies (\eg, cooling and power failures).
The system leverages historical temperature and power data, along with the adaptability of SaaS workloads, to:
(1) efficiently place new GPU workload VMs within cooling and power constraints,
(2) route LLM inference requests across SaaS VMs, and
(3) reconfigure SaaS VMs to manage load spikes and emergency situations.
Our evaluation on a large GPU cluster demonstrates significant reductions in thermal and power throttling events, boosting system efficiency.
\end{abstract}

\maketitle

\footnotetext[1]{Jovan Stojkovic is affiliated with the University of Illinois at Urbana-Champaign, but was at Microsoft Azure Research during this work.}

\section{Introduction}

\myparagraph{Motivation}
Generative large language models (LLMs) are increasingly used in various domains, such as healthcare~\cite{llmHealth}, developer productivity~\cite{copilot},
and education~\cite{education}.
This surge in usage drives high demand for LLM inference clusters~\cite{statChat}, requiring robust infrastructure with sophisticated software and costly hardware.
LLMs in the cloud typically run on virtual machines (VMs) powered by the latest GPUs, such as NVIDIA's A100~\cite{a100} and H100~\cite{h100}.
These GPUs consume significant power, 
challenging the cooling and power capacities of datacenters, major contributors to total cost of ownership (TCO)~\cite{overclockImmersion,cooldc,barroso2013warehouse}.
For instance, the A100 and H100 GPUs have thermal design powers (TDP) of 6.5 kW and 10.2 kW, 
and require substantial cooling capabilities to maintain safe operating temperatures.
Space constraints are typically less of an issue, as GPU servers and racks are power-dense.

Data centers hosting GPU workloads are organized into rows of server racks equipped with cooling systems to dissipate heat~\cite{coolprovision} and a power hierarchy for efficient power distribution~\cite{flexDatacenter}.
Cooling systems 
need to manage heat generated by the hottest server at any time.
However, cooling efficiency can vary spatially (\eg{}, some GPUs within a server may be hotter) and temporally (\eg{}, influenced by external temperatures).
Additionally, at each level of the power hierarchy, servers share a common power supply.
Hence, exceeding the total power draw leads to power capping.
Proper cooling and power provisioning is essential
during both normal operations and during cooling/power failures and emergencies.

While there have been advances in improving the performance efficiency of LLM inference clusters through software systems~\cite{orca,vllm,flashattention,spotserve,pets}, hardware techniques~\cite{splitwise,alizadeh2024llm,alisa}, and model architectures~\cite{llama3,falcon}, thermal and power  have not received the same attention~\cite{polca,towardsGreen,wordstowatts,vspetko2021dgx}.
Thermal~\cite{coolair,coolprovision,cooldc,coolEdge} and power~\cite{smartoclock,smoothOperator,flexDatacenter,thunderbolt,powerRouting} management of traditional datacenters have been extensively studied.
However, we observe that the unique characteristics of LLM inference workloads render the traditional proposals sub-optimal.

\begin{table}[t]
    \footnotesize
    \centering
    \begin{tabular}{ccccc}
        \toprule
        \textbf{Configuration parameters}                & Perf & Temp  & Power & Quality\\
        \midrule
        Model Size  (\eg{}, 70B$\rightarrow$7B)    & \greenuparrow & \greendownarrow & \greendownarrow & \reddownarrow\reddownarrow\\
        Quantization (\eg{}, FP16$\rightarrow$FP8) & \greenuparrow & \greendownarrow & \greendownarrow & \reddownarrow \\
        Parallelism (\eg{}, TP8$\rightarrow$TP2)   & \reddownarrow & \reduparrow     & \greendownarrow & \textcolor{black}{$-$} \\
        Frequency (\eg{}, 2GHz$\rightarrow$1GHz)   & \reddownarrow & \greendownarrow & \greendownarrow & \textcolor{black}{$-$} \\
        Batch Size (\eg{}, 64$\rightarrow$16)      & \reddownarrow & \greendownarrow & \greendownarrow & \textcolor{black}{$-$}\\
        \bottomrule
    \end{tabular}
    \caption{Impact on performance, temperature, power, and quality of an LLM inference server for multiple parameters.}
    \label{tab:design_space}
    \vspace{-9mm}
\end{table}

We target public clouds that host both Software-as-a-Service (SaaS) and Infrastructure-as-a-Service (IaaS) GPU workloads.
SaaS workloads are transparent GPU VMs managed by the cloud provider, while IaaS workloads are opaque GPU VMs with no provider visibility.
This allows configuration adjustments for SaaS workloads, while IaaS VMs remain unmodifiable.
The SaaS workload runs LLM inference, which involves several configuration parameters (\eg{}, GPU frequency, batch size, model parallelism, parameter count, and precision) that balance performance, thermal output, power, and result quality, as shown in \Cref{tab:design_space}.
In addition, LLM inference comprises distinct phases, each with unique performance, thermal, and power characteristics~\cite{splitwise}.

\myparagraph{Our work}
To address these challenges, we propose \emph{\system}, the first thermal- and power-aware scheduling scheme designed specifically for LLM inference clusters in the cloud.
\system{} maximizes cooling and power oversubscription while minimizing the impact on IaaS workloads and maintaining performance and accuracy for SaaS workloads.
In addition, \system{} dynamically adjusts LLM workloads in response to power or cooling failures in a datacenter.
The result is substantially reduced cloud platform TCO.

\system{} gracefully handles occasional load spikes and emergency events (\eg{}, cooling or power failures) through three core ideas.
First, it \emph{places GPU VMs} in a thermal- and power-aware manner by leveraging historical data on temperature, power consumption, and the load of a given service.
Second, it \emph{routes requests} across LLM instances based on the load of individual VMs, as well as the temperature and power slacks of the underlying infrastructure.
Third, it \emph{reconfigures} SaaS instances within their hierarchy until the temperature or power is reduced below safe limits.

\myparagraph{Results}
We evaluate \system{} on a large GPU cluster using production traces from a major cloud provider.
Our results show that \system{} maintains the P99 tail latency of inference requests while reducing maximum temperature by 17\% and peak row power by 23\%.
These reductions create more opportunities for oversubscription, enabling up to 40\% additional capacity and, consequently, lowering datacenter TCO.

To validate our findings at scale, we use traces from hundreds of production racks across a subset of datacenters and simulate \system{}.
Compared to other practical policies, \system{} reduces thermal and power throttling events by 97\% and 99\%, respectively.
In addition, we demonstrate that \system{} operates effectively during cooling and power failures.

\myparagraph{Summary}
We make the following main contributions:
\begin{itemize}[leftmargin=*]
    \item Characterization of thermal/power properties of GPU workloads and their behavior at production scale.
    \item \system{}, the first thermal- and power-aware scheduling scheme for LLM inference systems.
    \item A thorough evaluation of \system{} in a GPU cluster using large-scale production traces.
\end{itemize}

\section{Characterizing Challenges in Thermal and Power Infrastructure for GPUs}
\label{sec:characterization-infra}

To identify the challenges in managing cooling and power for GPU workloads, we characterize the datacenter infrastructure required to support these workloads.
We focus on spatial and temporal heterogeneity in the usage of thermal and power infrastructure, that can be exploited to operate GPU workloads more efficiently.
We introduce equations to help us model thermal and power aspects at datacenter scale.

\myparagraph{Datacenter overview}
Cloud providers host a variety of services from multiple users on shared infrastructure.
We study datacenters hosting both A100~\cite{a100} and H100~\cite{h100} GPUs, which are typically used for LLMs~\cite{splitwise}.
Servers in a datacenter are arranged in rows of racks.
Due to the size and power density of GPUs,
racks and rows host fewer servers than in general-purpose datacenters.
These datacenters also host other infrastructure for storage, network, and management.

\subsection{Cooling}

\myparagraph{Infrastructure}
GPU servers generate a large amount of heat, while their temperature needs to stay below a specific threshold (\eg, 85\textdegree C for GPUs).
On exceeding the threshold, the hardware starts throttling the computation to prevent 
failures and permanent damage~\cite{h100-cooling}.

Depending on the regional climate, datacenters may use technologies like mechanical or adiabatic cooling to lower temperatures~\cite{daraghmeh2017review}.
While other alternatives exist (\eg, liquid cooling~\cite{overclockImmersion}), we focus on air cooling as it is the most commonly used method in today’s datacenters~\cite{meta-sustainability,energygov-evaporative,daraghmeh2017review,google-cooling}.
Many of our insights can be applied to other technologies.

\begin{figure}[t!]
    \centering
    \includegraphics[width=0.95\linewidth]{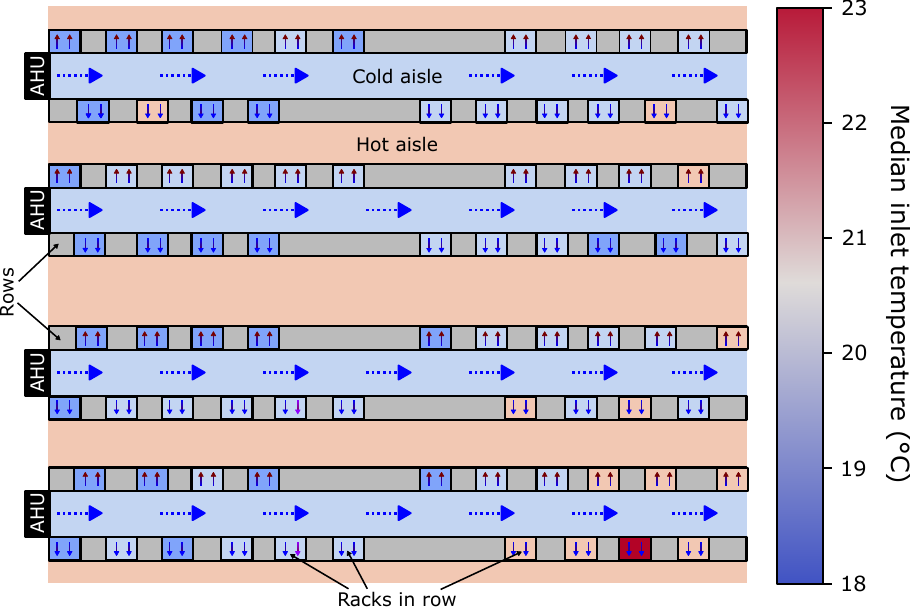}
    \caption{
    Sample datacenter layout illustrating 80 racks organized into 8 rows with 4 cold aisles.
    The rack color represents the inlet temperatures for the top server.
    }
    \label{fig:hot_cold_aisle}
    \vspace{-2mm}
\end{figure}

Datacenters are usually arranged in aisles composed of two rows.
\Cref{fig:hot_cold_aisle} illustrates an example of airflow within one of the rooms in one of the datacenters.
The air handling units (AHUs) in each row blow cold air from the datacenter-level cooling devices (\eg, adiabatic cooling towers in evaporative cooling) into the cold aisle which is contained.
The servers use fans with modulated speeds based on activity to draw cold air from the front, pass it through the server (including the GPUs), and exhaust the heated air into the hot aisle.
The cooling devices then take this hot air and cool it down again.
To avoid heat recirculation (\ie, hot air returning to the cold aisle), the airflow provided by the AHUs must exceed the airflow consumed by the servers in the cold aisle.

\myparagraph{Provisioning}
Datacenter operators usually provision the cooling infrastructure to sustain their peak load~\cite{coolprovision}.
This means they need to have (1) enough airflow in each aisle (\ie, AHUs) to prevent heat recirculation and (2) enough cooling capacity in the datacenter to lower the temperature within operating conditions.
Operators can add racks to rows as long as both conditions are met.

\myparagraph{Failures}
Datacenters typically build redundancy to handle failures (\eg{}, N+1~\cite{flexDatacenter}).
When a cooling device fails, other devices in the datacenter compensate.
However, this results in reduced cooling capacity, which may increase the temperature across all cold aisles in the datacenter.
If an AHU fails, the other AHUs in the aisle handle the airflow.
Insufficient airflow from the AHUs in the aisle leads to heat recirculation, raising the temperature of all servers in the two rows.

\myparagraph{Characterization}
To understand the thermal impact of the cooling infrastructure, we study a sample of our datacenters at \provider{} containing tens of thousands of GPUs (including both A100 and H100) across three regions with varying climates.
The study spans three months, from June to October 2024, covering the warmer months in these locations.
We collect data on inlet and outlet temperatures for each server, the outside temperature, and the temperature and power of each component (\eg{}, GPU and  memory), reporting the averages every 10 minutes.
This 10-minute interval aligns with the frequency of all sensors and enables the approximation of heat from average power.
While we discuss individual examples, our insights are derived from the full dataset.

\begin{figure}
    \centering
    \includegraphics[width=\linewidth]{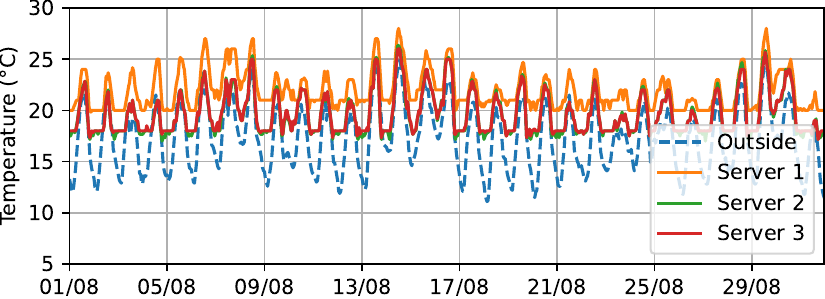}
    \caption{Inlet and outside temperatures for three servers throughout August 2024.}
    \label{fig:char_inlet_outside}
    \vspace{-3mm}
\end{figure}

\begin{figure}
    \centering
    \includegraphics[width=\columnwidth]{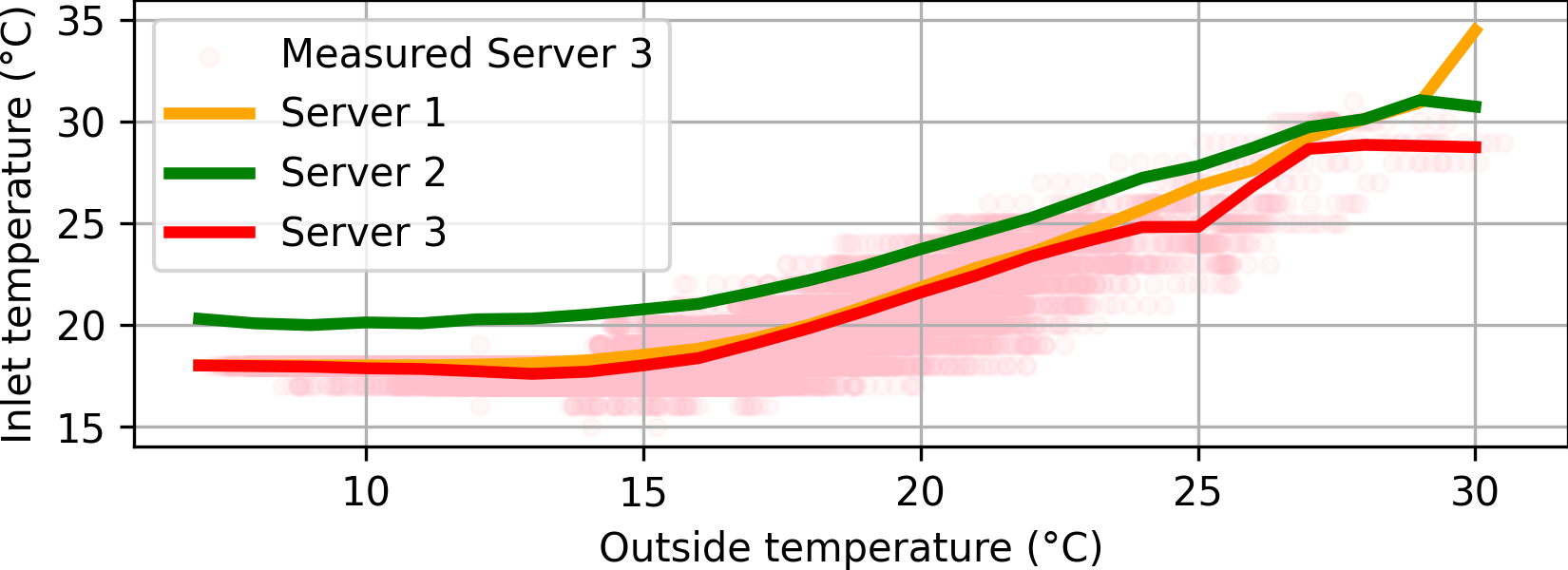}
    \caption{
    Regression analysis comparing inlet and outside temperatures for three sample servers.
    Includes actual measurements for Server 3.
    }
    \label{fig:char_inlet_outside_regression}
    \vspace{-4mm}
\end{figure}

\myparagraphemph{Outside temperature}
Cooling technologies like adiabatic cooling~\cite{coolprovision,coolair,meta-sustainability,energygov-evaporative,google-cooling,daraghmeh2017review} use outside air when it is cold for efficiency.
\Cref{fig:char_inlet_outside} shows the inlet temperature for three servers in the same aisle and the outside temperature in August 2024.
The inlet temperature follows the trend of the outside temperature.
\Cref{fig:char_inlet_outside_regression} shows the inlet and outside temperature for these servers over three months.
Each point represents the inlet temperature for Server 3 and the outside temperature every 10 minutes.
The lines are a regression of these points for each of the servers.
When it is cold outside, the cooling maintains the inlet temperature 
(\eg{}, over 18\textdegree{}C) to avoid increasing humidity which increases failures~\cite{coolair}.
After 15\textdegree C outside, the inlet temperature increases linearly with the outside.
When it is hot outside (\ie, 25\textdegree{}C), the cooling lowers the temperature further.
Locations with higher temperatures are less sensitive to the outside temperature.

\begin{figure}[t!]
    \centering
    \includegraphics[width=\columnwidth]{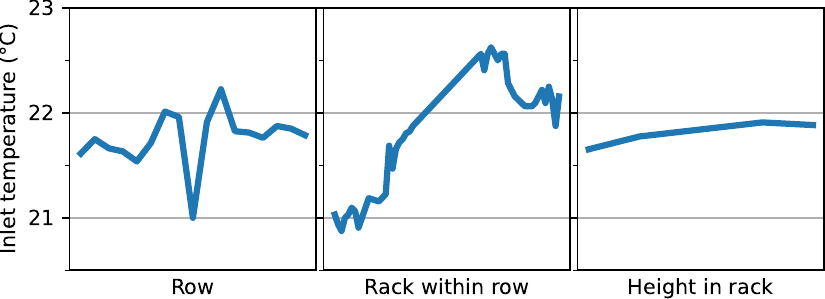}
    \caption{Inlet temperature distribution across physical entities:
    rows, racks within rows, and height within racks.
    }
    \label{fig:char_inlet_physical_entity}
\end{figure}

\myparagraphemph{Datacenter layout}
Given the physical layout, the inlet temperature for each server is not homogeneous, and there are hotspots.
\Cref{fig:char_inlet_outside,fig:char_inlet_outside_regression} show how Server 2 is consistently warmer ($\sim 2$\textdegree{}C) than the other two.
\Cref{fig:hot_cold_aisle} shows that the median temperature across racks and rows varies: the end of some rows is warmer than others because of airflow and construction differences.
These airflow patterns are hard to estimate beforehand and require measuring them empirically or expensive simulations.
\Cref{fig:char_inlet_physical_entity} shows the median temperature over the three months depending on the physical entity.
Some rows have temperatures up to 1\textdegree{}C higher than others, with racks within a row showing differences of up to 2\textdegree{}C.
The height within the rack has a minor impact.

\begin{figure}[t!]
    \centering
    \includegraphics[width=\columnwidth]{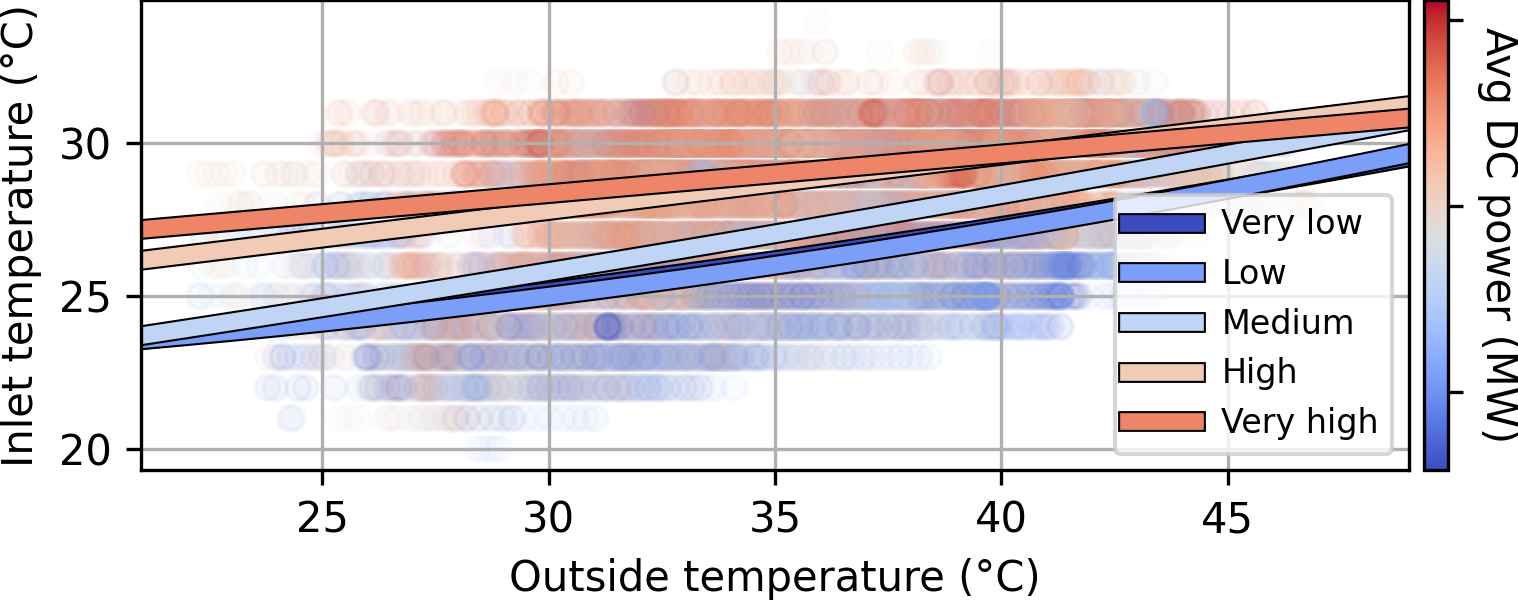}
    \caption{
    Inlet temperature as a function of datacenter load and outside temperature.
    It includes actual measurements and regression lines per power load levels.
    }
    \label{fig:char_inlet_load}
\end{figure}

\myparagraphemph{Datacenter load}
The amount of heat in the datacenter also affects the inlet temperature.
Cooling devices usually lower the outlet temperature by $\Delta T$ (\eg, 10\textdegree{}C).
When the heat generated by the servers increases, the inlet temperature also increases.
\Cref{fig:char_inlet_load} shows the regression between datacenter load (average power for 10 minutes) and the difference in inlet temperature for one server in a hot region.
For example, when it is 35\textdegree{}C outside, there is an inlet temperature difference of 2\textdegree{}C when the load is low and high.
Note that the correlation with datacenter load is much lower than with inlet temperature.
Using the three months of data, we apply regression to model the inlet temperature for each server $s$:
\begin{equation}
\forall_{s \in S} T_{inlet,s} = f_{inlet,s}(T_{outside}, \text{Load}_{\text{DC}})
\label{eq:t_inlet}
\end{equation}

\begin{figure}
    \centering
    \includegraphics[width=\columnwidth]{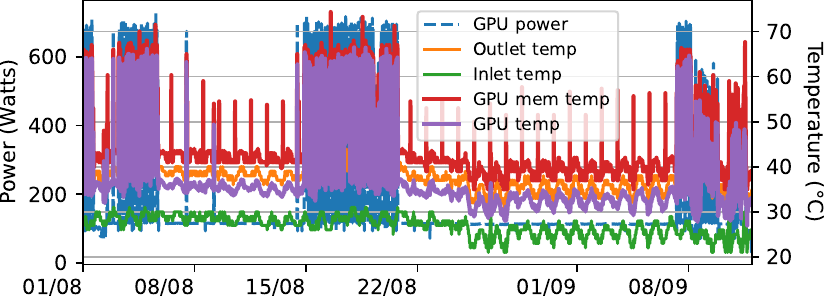}
    \caption{
    GPU temperature and over time alongside inlet temperature.
    }
    \label{fig:char_gpu_inlet_load}
    \vspace{-3mm}
\end{figure}

\begin{figure}
    \centering
    \includegraphics[width=\columnwidth]{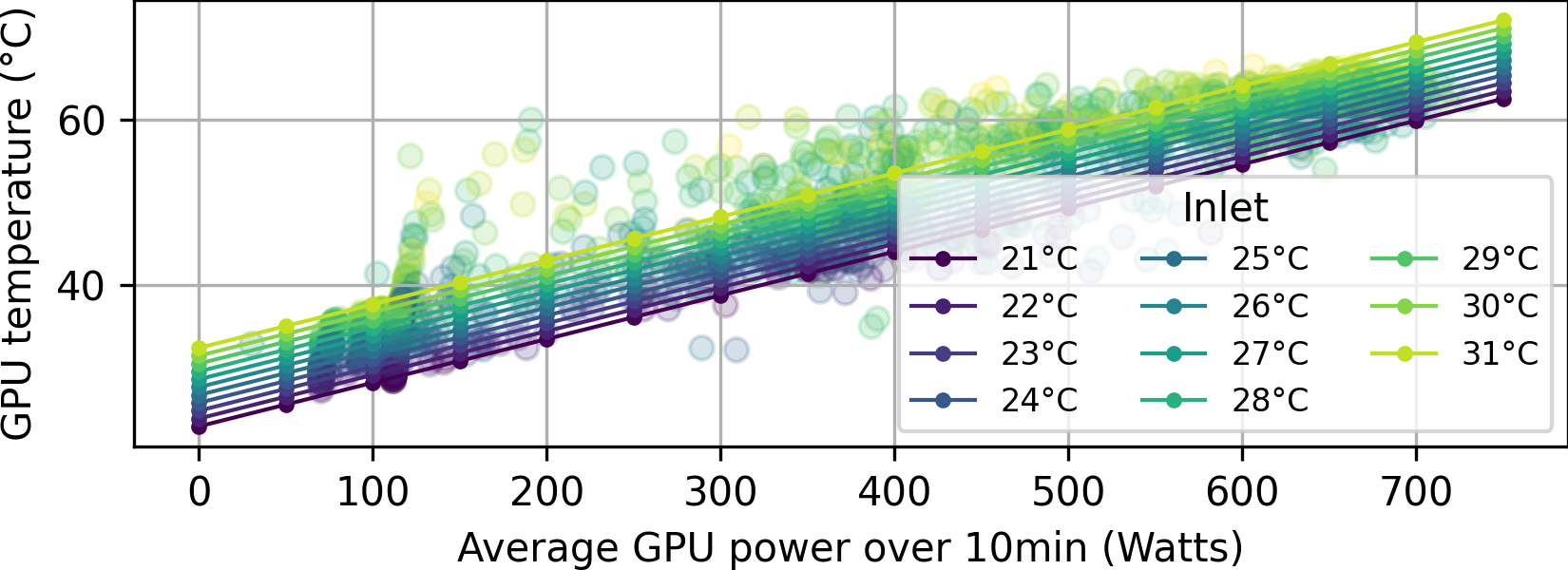}
    \caption{
    GPU temperature as a function of inlet temperature and GPU power load over 10 minutes for \Cref{fig:char_gpu_inlet_load}.
    Includes a regression curve based on inlet temperature and GPU load.
    }
    \label{fig:char_gpu_inlet_load_regression}
    \vspace{-4mm}
\end{figure}

\myparagraphemph{GPU temperature}
Once the inlet air enters the server, the fans circulate the cold air to cool down the server components (\eg{}, GPUs and their memory).
\Cref{fig:char_gpu_inlet_load} shows the GPU and memory temperature, along with the inlet and outlet temperature, and GPU power for an example server running tests for this work over
45 days.
The GPU memory is warmer than the GPU and there is an offset between inlet and outlet.
\Cref{fig:char_gpu_inlet_load_regression} displays a linear regression of the temperature of one GPU compared to the inlet temperature and GPU load.
This regression has a mean absolute error of less than 1\textdegree{}C.
The GPU temperature is sensitive to both the GPU load and the server inlet temperature.
This regression also captures the inlet temperature increase caused by power leakage~\cite{aguilera2014process}.

\begin{figure}
    \centering
    \includegraphics[width=\columnwidth]{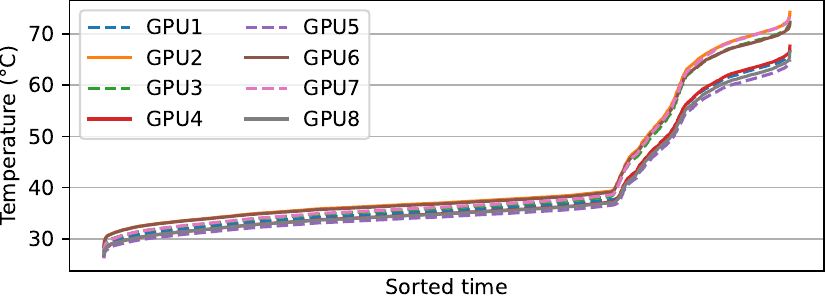}
    \caption{Sorted GPU temperatures for the 8 GPUs corresponding to the data in \Cref{fig:char_gpu_inlet_load}.}
    \label{fig:char_8_gputemp}
    \vspace{-1.5mm}
\end{figure}

\myparagraphemph{GPU heterogeneity}
\Cref{fig:char_8_gputemp} shows the temperatures of the 8 GPUs in a DGX A100~\cite{a100} server running the same workload.
Despite identical inlet temperatures and GPU utilization at each point in time, the temperatures of individual GPUs can differ by up to 10\textdegree{}C.
This variation is due to server layout (\eg{}, components obstructing airflow to certain GPUs) and manufacturing variations in the GPUs themselves (\ie, process variation~\cite{aguilera2014process}).
When the temperature within the server is high, the fans will increase airflow to cool the GPUs.
It is important to account for this when provisioning the AHUs.

\begin{figure}
    \centering
    \includegraphics[width=\columnwidth]{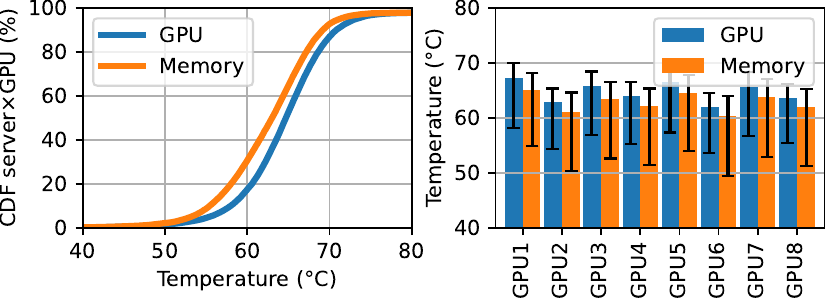}
    \caption{Distribution of temperatures for over 3,000 GPUs and their memory at high load with comparable inlet temperatures in a single data center.}
    \label{fig:char_gputemp_dc_cdf}
    \vspace{-4mm}
\end{figure}

\Cref{fig:char_gputemp_dc_cdf} shows the heterogeneity in temperature across GPUs in one datacenter with high GPU load and similar inlet temperatures.
There is a range of over 20\textdegree{}C across GPUs in the same datacenter.
The temperature of the GPU memory is slightly lower than the GPU itself.
The right side of \Cref{fig:char_gputemp_dc_cdf} shows the median temperature for each of the 8 GPUs in the server and their inter-quartile range.
The GPUs with even identifiers (\eg, GPU2 and GPU4) have a lower temperature due to the server layout, 
as they are closer to the inlet~\cite{vspetko2021dgx}.

Using the data for the three months, we generate a model for the temperature of each GPU $g$ in each server $s$:
\begin{equation}
\forall_{s \in S, g \in G} T_{\text{GPU } s,g} = f_{GPU s,g}(T_{inlet,s}, \text{Load}_{\text{GPU },g})
\label{eq:t_gpu}
\end{equation}

\myparagraphemph{Airflow}
We measure the speed of the server fans when the server is idle and when it is running at full load (\ie{}, all GPUs running heavy workloads).
Then we run a few intermediate settings and interpolate a linear function.
Our measurements match the manufacturer specs, which indicate an airflow of 840 and 1105 cubic feet per minute (CFM) at 80\% speed with pulse width modulation (PWM) fans for A100 and H100 respectively~\cite{a100,h100}.
All servers follow a similar linear function with very small differences: $f_{air}(Load_{GPU,s})$.
As mentioned, we need to guarantee the provisioned airflow from the AHU is larger than the aggregated server airflow requirement:
\begin{equation}
\forall_{\text{Aisle} \in DC} \sum_{s \in S_{\text{Aisle}}} f_{\text{air}}(\text{Load}_{GPU,s}) \le \text{ProvAHU}_{\text{Aisle}}
\label{eq:cfm}
\end{equation}

\myinsight{}
For effective thermal management, cloud operators must consider temporal heterogeneity due to variations in outside temperature and load, spatial heterogeneity related to datacenter and server layouts, and airflow requirements.

\subsection{Power}

\myparagraph{Infrastructure}
Datacenters typically implement a three-level power distribution hierarchy to deliver electricity from the utility grid to individual servers~\cite{wang2009ship, flexDatacenter, wu2016dynamo}.
At the highest level, an Automatic Transmission Switch (ATS) directs power from the grid to multiple Uninterruptible Power Supplies (UPS).
Each UPS shares a fraction of the total datacenter power load and is connected to a series of Power Distribution Unit (PDU) pairs.
These PDU pairs further step down the voltage and support multiple rows of server racks. 

\myparagraph{Provisioning}
To prevent tripping the circuit breakers, datacenter operators provision for peak power usage at each level of the hierarchy to account for worst-case scenarios. 
For safety, when the total power draw exceeds the power supply, servers within that level are power-capped.
For design simplicity and to reduce implementation costs, these power limits are further distributed down the hierarchy homogeneously, eventually limiting the number of server racks that can be provisioned within the budget.
Operators can oversubscribe the power capacity by adding racks to a row, as long as they remain within the row-level power envelope.

\myparagraph{Failures}
To ensure high availability, clouds implement redundancy at each power hierarchy level.
For instance, prior work describes a setup with 4N/3 redundancy at the UPS level and 2N at the PDU level~\cite{flexDatacenter}.
When a UPS fails, its load is redistributed among the remaining three units.
Under heavy load, this can push the others over capacity, requiring each unit to quickly reduce its load to maintain limits, effectively lowering the datacenter capacity to 75\%.
We focus on this design as its balances normal and fail-over operation.
Our findings extend to other redundancy models~\cite{google-mvpp, capmaestro}.

\myparagraph{Characterization}
Prior works characterize the power profile of GPU servers running LLMs~\cite{polca, splitwise}.
We complement these studies with our own data focusing on power imbalances at datacenter scale.
We study the same datacenters as in the cooling section.
For confidentiality, we normalize all values to the maximum power draw.

\myparagraphemph{GPU load and server power}
We measure server power for both A100 and H100 servers across various utilization levels and workloads (\eg{}, \Cref{fig:char_gpu_inlet_load}).
Server power is strongly correlated with GPU load~\cite{polca,vspetko2021dgx}.
Even when idle, servers consume significant power, similar to traditional CPU servers~\cite{meisner2009powernap}.
Besides GPU power, a substantial portion is drawn by fans, storage, memory, CPUs, and other components~\cite{polca}.
For each server $s$, we used polynomial regression to generate a function $f_{power}(\text{Load}_{\text{GPU } g,s})$, which accounts for fan power and other components that also depend on load.

\begin{figure}[t!]
\subfloat[Power utilization over 1-week.]{
  \includegraphics[clip,width=0.5\columnwidth]{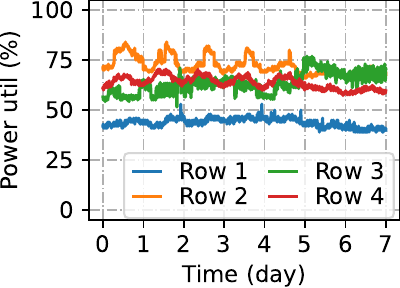}
  \label{fig:power_row}
}
\subfloat[Row power distribution.]{
  \includegraphics[clip,width=0.5\columnwidth]{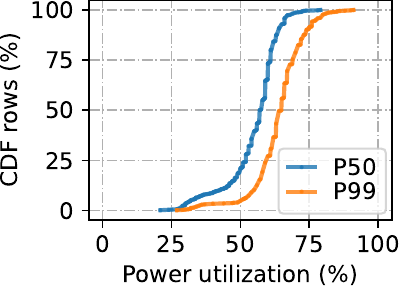}
  \label{fig:power_dist}
}
\vspace{-3mm}
\caption{
Power utilization timeline for four sample rows and power across rows within a datacenter.}
\label{fig:power_time}
\end{figure}

\myparagraphemph{Power imbalance across rows}
Row power utilization is the aggregation and multiplexing of individual server power.
\Cref{fig:power_time} shows the power draw of four sample rows in a datacenter over a week.
While most rows exhibit lower power draw, a few have significantly higher consumption.
We quantify this behavior at scale in \Cref{fig:power_time}, which shows the CDF of P50 and P99 power draw across 100 rows in a subset of our datacenters.
The figure shows a \emph{heavy tail} pattern: 50\%, 75\%, and 90\% of the rows draw 28\%, 18\%, and 10\% less P99 power than the most power-hungry row, respectively.

The high-power rows create hotspots in the datacenter, requiring sufficient power provisioning to meet the demands of these rows.
Thus, power allocated to lower-demand rows is wasted, significantly hindering the cloud provider's ability to safely oversubscribe.
We define this as:
\begin{equation}
\forall_{\text{Row} \in DC} \sum_{s \in S_{\text{Row}}} Power_{s}(\text{Load}_{GPU,s}) \leq \text{ProvPower}_{\text{Row}}
\label{eq:power}
\end{equation}

\myinsight{}
The power demands of GPU clusters present a strong opportunity for power oversubscription.
However, to safely oversubscribe, the infrastructure must effectively manage the rows at the tail end that generate hotspots.

\section{Characterizing Opportunities in Thermal and Power Properties of GPU Workloads}
\label{sec:characterization}

To reason about the impact of GPU workloads on thermal and power properties in cloud datacenters, we
(1) analyze the physical placement of GPU workloads at \provider{},
(2) profile SaaS LLM inference workloads from a production environment, and
(3) characterize the thermal and power properties of LLM inference varying a set of configurations with open-source models~\cite{llama2}.

\begin{figure}[t]
    \centering
    \subfloat[VM placement.]{%
        \includegraphics[clip,width=0.6\columnwidth]{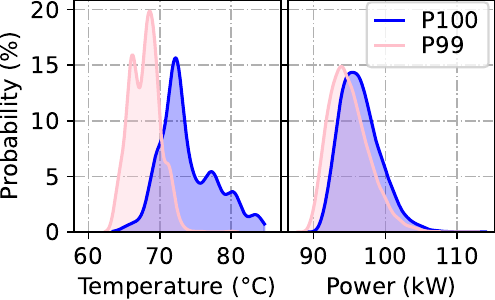}
        \label{fig:char_placement_power}
    }
    \subfloat[Temperature vs. power.]{
        \includegraphics[clip,width=0.4\columnwidth]{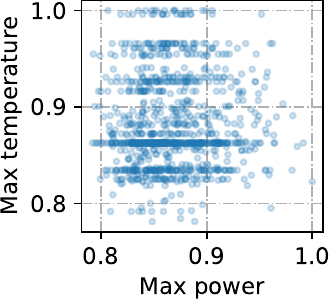}
        \label{fig:placement_temp_vs_power}
    }
    \vspace{-3mm}
    \caption{Distribution of aisle peak GPU temperature and row power with 100K random VM placements.
    }
    \label{fig:char_placement}
\end{figure}

\subsection{GPU workload placement}

\myparagraph{Cooling and power impact}
To analyze the impact of GPU workloads on cooling and power infrastructure, we deploy 80 VMs across two rows in a datacenter (\Cref{sec:characterization-infra}) and generate 100K random placements across these two rows.
\Cref{fig:char_placement} illustrates the distribution of peak server temperatures and row power with these placements.
The worst-case placement results in a maximum temperature exceeding 85\textdegree{}C, while a typical placement averages around 72\textdegree{}C.
In terms of peak power, the worst-case placement increases peak power by 27\% over the optimal placement.
If more intensive workloads are placed on hotter servers or if synchronous peak loads are co-located on the same row, the provider must provision sufficient cooling and power to support these extreme scenarios.
Additionally, \Cref{fig:placement_temp_vs_power} shows no correlation between maximum temperature and peak power for VM placements, indicating that cloud providers should consider both thermal and power factors when placing VMs across the datacenter.

\begin{figure}[t]
    \centering
    \subfloat[IaaS and SaaS VMs lifetimes.]{%
        \includegraphics[clip,width=0.5\columnwidth]{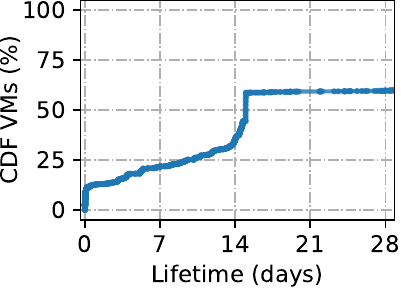}
        \label{fig:char_lifetime}
    }
    \subfloat[VMs per SaaS endpoint.]{%
        \includegraphics[clip,width=0.5\columnwidth]{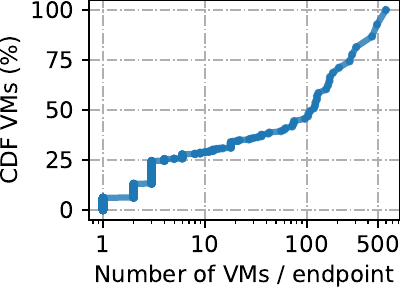}
        \label{fig:char_servicedist}
    }
    \vspace{-3mm}
    \caption{VM lifetime and number of VMs per endpoint.}
\end{figure}

\begin{figure}[t]
    \centering
    \subfloat[VM load.]{%
    \includegraphics[width=0.5\columnwidth]{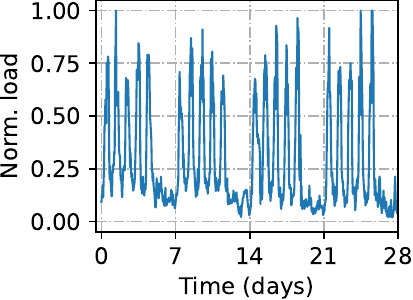}
    \label{fig:char_periodic_vm}}
    \subfloat[Row power.]{%
    \includegraphics[width=0.5\columnwidth]{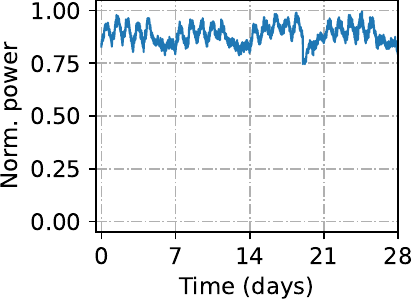}
    \label{fig:char_periodic_row}}
    \vspace{-3mm}
    \caption{Normalized load over time for an example VM and power over time for an example row.}
    \label{fig:char_periodic}
\end{figure}

\myparagraph{Long-lived VMs}
\Cref{fig:char_lifetime} shows the lifetime for VMs running GPU workloads.
Most VMs are long-lived (\eg{}, over 60\% run for more than two weeks).
Since these VMs occupy a full server, this implies that a given server may be dedicated to a workload for an extended period.
\Cref{fig:char_periodic_vm} shows an example VM over a 4-week period with a distinctly periodic diurnal load pattern.
Aggregated at row level (\Cref{fig:char_periodic_row}), the power consumption also shows periodic pattern.

\begin{figure}[t]
    \centering
    \subfloat[Row-based.]{%
      \includegraphics[clip,width=0.5\columnwidth]{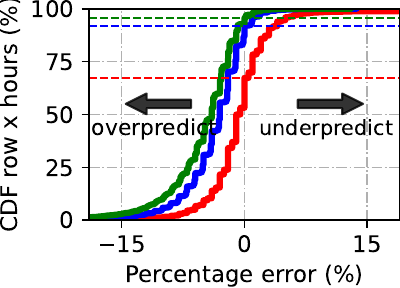}
      \label{fig:row_accuracy}
    }
    \subfloat[Customer-based.]{%
      \includegraphics[clip,width=0.5\columnwidth]{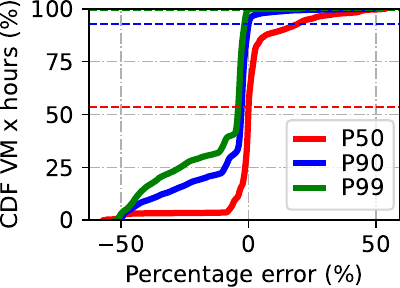}
      \label{fig:sub_accuracy}
    }
    \vspace{-3mm}
    \caption{CDF of the row- and customer-based power draw prediction error with different power templates.}
    \label{fig:eval_accuracy}
\end{figure}

\myparagraph{Predictable load}
\Cref{fig:row_accuracy} shows that using different power templates\cite{smartoclock}, row power prediction based on past history has less than 10\% error for most row hours.
A conservative prediction using P99 underestimates the power for less than 4\% of the row$\times$hours.
For VM power prediction, cloud providers can leverage customer information, as shown in \Cref{fig:sub_accuracy}, with errors below 10\% for more than 75\% VM$\times$hours and underpredictions between 2-7\% for P90 and P99 templates.
These further demonstrate the predictability of row- and VM-level power consumption.

\myinsight{}
Cloud operators can leverage workload heterogeneity and predictability for intelligent workload placement to relieve hotspots and smooth out thermal/power spikes.

\subsection{LLM inference routing}

\myparagraph{IaaS vs SaaS}
\provider{} hosts a variety of GPU workloads, which we categorize categorized as either IaaS or SaaS.
IaaS uses opaque VMs~\cite{azure-gpu,ec2-p4,ec2-p5,gcp-gpu} and customers can run any workload (\eg, inference, training, fine-tuning) for any model (\eg, LLM, diffusion, image recognition).
On the other hand, SaaS is fully managed by the cloud provider~\cite{azure-ml,gcp-gemini,aws-sagemaker,aws-sagemaker-llama}.
In all the datacenters across regions, there is a significant portion of SaaS VMs.
This allows for more flexible thermal and power shaping in the datacenter.

\myparagraph{SaaS workloads}
The SaaS offering in our A100 and H100 datacenters serves multiple LLM inference endpoints, each hosting various LLMs for different applications~\cite{ms-endpoint,hf-endpoint,gcp-gemini,aws-sagemaker}.
Each endpoint operates a dedicated set of VMs, which can host multiple LLM inference instances, routing user inference requests across these instances.
\Cref{fig:char_servicedist} shows the distribution of VMs serving requests per SaaS inference endpoint:
50\% of the VMs belong to large endpoints with over 100 VMs spanning across multiple rows.

\myparagraph{Load balancing}
Our SaaS implementation distributes LLM inference requests across VMs to enhance latency and throughput~\cite{sarathi-serve,li2024llm,yu2023stateful}.
However, these VMs may be placed in rows with different thermal and power characteristics (\eg{}, \Cref{fig:power_time}).
If the Load Balancer is unaware of temperature conditions, it may assign equal workloads to servers that are already near thermal throttling.
Similarly, disregarding power conditions could result in sending additional load to VMs in rows with high power demand from neighboring IaaS servers, exacerbating power strain.

\myinsight{}
Cloud operators can leverage the flexibility of SaaS LLM inference workloads for thermal- and power-aware request routing to maximize oversubscription opportunities.

\subsection{LLM inference instance configuration}
\label{sec:char_knobs}

As outlined in \Cref{tab:design_space}, LLM inference servers have various configuration options that balance thermal/power requirements, performance, and result quality.
To evaluate these options, we run Llama2~\cite{llama2} inference workloads on an NVIDIA DGX A100 server~\cite{a100}.
LLM inference consists of two distinct phases~\cite{splitwise}:
the prefill phase, which processes the entire prompt in parallel, and the decode phase, which generates each output token sequentially.

\begin{figure}[t]
    \centering
    \subfloat[Tensor parallelism~\cite{shoeybi2019megatron}.]{%
        \includegraphics[width=\columnwidth]{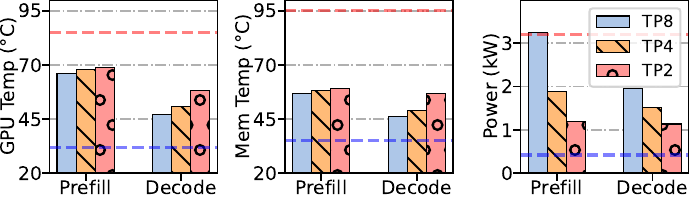}
        \label{fig:char_parallelism}
    }
    \vspace{2mm}
    \subfloat[Batch size~\cite{orca}.]{%
        \includegraphics[width=\columnwidth]{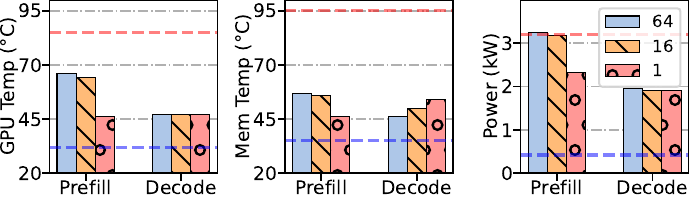}
        \label{fig:char_batchsize}
    }
    \vspace{2mm}
    \subfloat[Llama2 model sizes~\cite{llama2}.]{%
        \includegraphics[width=\columnwidth]{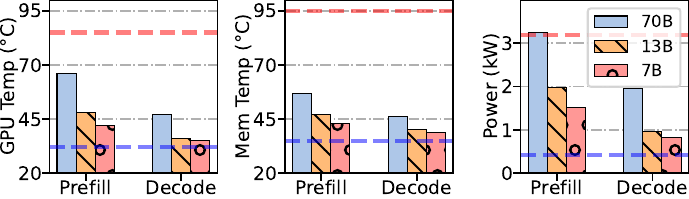}
        \label{fig:char_modelsize}
    }
    \caption{Temperature and power of a server running prefill and decode phases varying configuration parameters.}
    \label{fig:char_knobs}
\end{figure}

\myparagraph{Impact of configuration parameters}
In \Cref{fig:char_knobs}, we quantify the impact of each parameter individually:
GPU frequency, parallelism, batch size, model size, and quantization.
The red line represents the maximum temperature and power (TDP) while the blue line shows the idle.

\myparagraphemph{GPU frequency}
Lowering the frequency reduces both temperature and power of the GPU.
Prompt phases are more sensitive to GPU frequency~\cite{polca, towardsGreen}.
Although reducing the frequency has lower impact on temperature and power than other configuration parameters, it does not affect the quality of results and can be applied instantaneously due to its relatively low overhead.

\myparagraphemph{Model parallelism}
We focus on tensor parallelism~\cite{shoeybi2019megatron} because other parallelisms like data and pipeline parallelism are not as effective for LLM inference within a single server~\cite{splitwise}.
\Cref{fig:char_parallelism} shows the temperature and power of a server running prompt and decode phases varying tensor parallelisms: TP8, TP4, and TP2~\cite{shoeybi2019megatron} (\ie{}, powers of two with the number of KV heads~\cite{llama2}).
With TP2, the total server power reduces as we use less GPUs.
However, as the same amount of work is concentrated in fewer GPUs, the per-GPU power increases.
Hence, the temperature of the hottest GPU increases.
Lower parallelism impacts power more significantly during prompt phase and temperature during decode phase.
Thus, depending on the workload's phase and the target metric (\ie{}, reducing temperature or power), the system needs to take different actions.

\myparagraphemph{Batch size}
\Cref{fig:char_batchsize} shows the temperature and power with different batch sizes: 64, 16 and 1~\cite{orca}.
Due to the reduced computational intensity, temperature and power reduce.
However, during decode phase, as GPUs need to more frequently fetch the data from memory via the memory controller (instead of bulk transfers), the memory temperature increases.
Thus, depending on the bottleneck (temperature or power) and the workload's phase (prompt or decode), the system may decide to choose different batch sizes.

\myparagraphemph{Model size}
\Cref{fig:char_modelsize} shows the power consumption and temperature associated with different Llama2~\cite{llama2} model sizes: 70B, 13B, and 7B.
As the model size decreases, the computational intensity of inference reduces significantly, resulting in lower power draw and temperature.
However, smaller models tend to produce results of lower quality~\cite{qserve}.
For example, the 7B model reduces result quality by 30-40\% compared to the 70B model~\cite{badri2023hqq, llama2}.

\myparagraphemph{Model quantization}
We observe a similar behavior with quantized model: lower precision leads to reduced temperature and power while incurring 2-20\% accuracy impacts~\cite{qserve, badri2023hqq, li2024evaluating}.
Because both smaller and quantized models generally lead to reduced quality, the system must carefully manage the proportion of the load directed to different model variants to uphold average per-service quality SLOs.

\begin{figure}[t]
    \centering
    \includegraphics[width=\columnwidth]{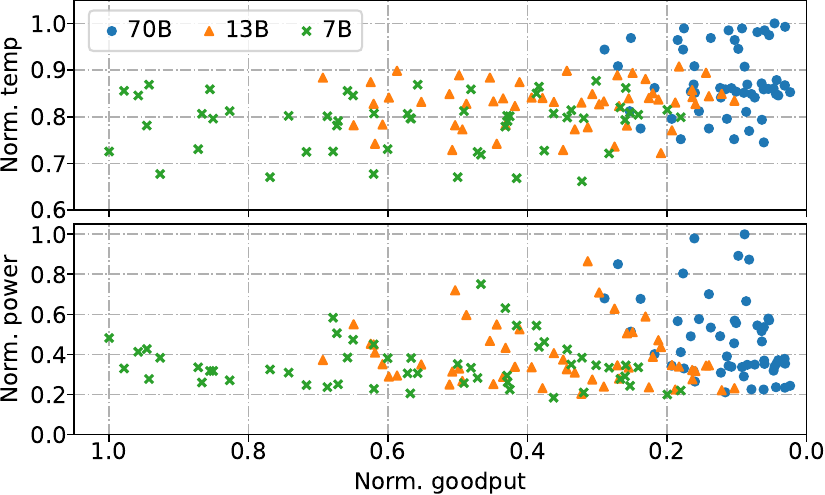}
    \caption{
    Normalized temperature and power (lower is better) and goodput (higher is better) of Llama2~\cite{llama2} with all configuration parameters highlighting the model size.
    }
    \label{fig:temp_power_perf_3d}
\end{figure}

\myparagraph{Thermal and power space}
We quantify the performance of an LLM inference server in terms of its \emph{goodput} (\ie{}, the number of tokens processed per second while staying within TTFT and TBT SLOs, defined as 5$\times$ the execution time on an unloaded system~\cite{alpaserve,zhong2024distserve}).
\Cref{fig:temp_power_perf_3d} illustrates the trade-off between temperature/power and goodput across all configurations (\ie, GPU frequency, parallelism, batch size, model size, and quantization).
The figure highlights the impact of the model size as it affects quality.
Each model has a Pareto frontier representing configurations that minimize temperature and power  with minimal impact on performance.

\myinsight{}
Cloud operators can effectively shape thermal and power while minimally impacting LLM inference performance and results quality by configuring SaaS instances.

\section{\system{} Design}
\label{sec:design}

\myparagraph{Architecture}
Based on our insights, we propose \system{}, a framework for thermal and power management in GPU clusters for cloud environments.
\system{} is specifically designed to address the unique properties and challenges of LLM inference workloads.
\system{} enhances cooling and power oversubscription capabilities while effectively managing infrastructure failures, thereby reducing datacenter TCO with minimal impact on workload performance or result quality.
The system leverages the adaptability of SaaS workloads without impacting IaaS workloads.

\begin{figure}[t]
    \centering
    \includegraphics[width=\columnwidth]{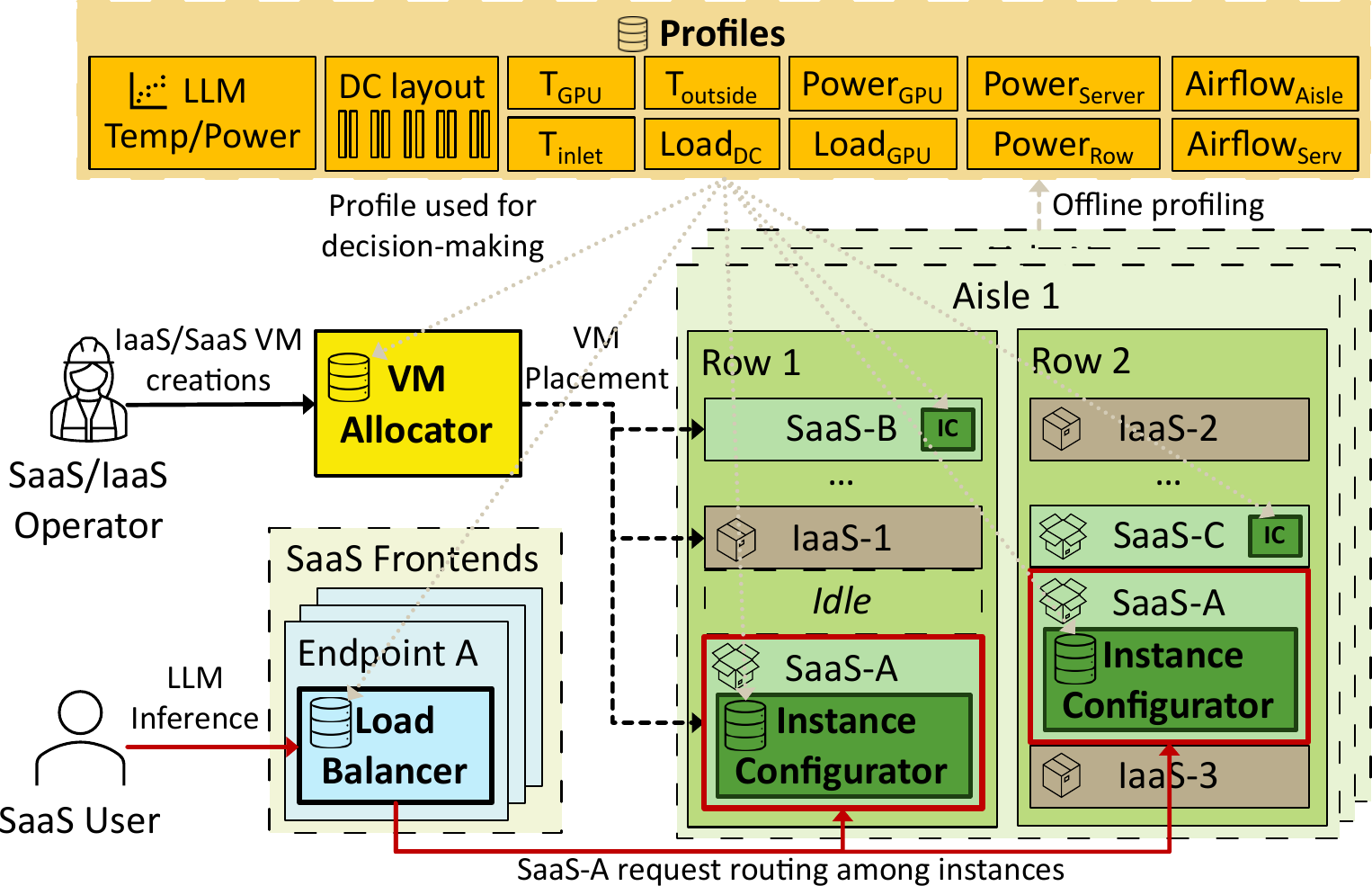}
    \caption{\system{} architecture overview.}
    \label{fig:arch}
\end{figure}

\Cref{fig:arch} overviews the architecture.
\system{}
(1) extends existing components of conventional cloud LLM inference clusters (\eg{}, per-cluster \emph{VM Allocator} and per-endpoint \emph{Load Balancer}),
(2) introduces a per-SaaS VM \emph{Instance Configurator}, and
(3) maintains multiple \emph{Profiles}.
To achieve its goals, \system{} focuses on three core aspects:
VM placement, LLM inference request routing, and instance configuration.

\subsection{Workload placement}
\label{sec:design-placement}

As VMs arrive, the \emph{VM Allocator} assigns each new VM (whether IaaS or SaaS) to a server, aiming to meet workload demands while minimizing the risk of thermal or power hotspots.

\myparagraph{Aisle and row filtering}
Using historical server load data, we predict peak the airflow requirements for each aisle and peak power demand for each row.
If data is insufficient for a server (\ie{}, less than one week), we assume peak load conditions.
We then estimate the load of the new VM based on the load from VMs associated with
(1) the same user for IaaS workloads and
(2) the same endpoint for SaaS workloads.
Again, we assume peak load if historical data is insufficient.
We estimate the new peak airflow for each aisle and power demand for each row if the VM were to be placed, and we filter out servers in aisles or rows that would exceed airflow or power limits (\Cref{eq:cfm,eq:power}).

\myparagraph{Placing hotter IaaS VMs in cooler servers}
As fine-grained control over IaaS VMs is limited, we aim to place these VMs in cooler servers.
For each server, we feed the historical server inlet temperature and the predicted VM load into \Cref{eq:t_gpu} to estimate peak GPU temperature.
We then select the servers with the lowest projected temperatures.
We attempt to place SaaS VMs in warmer servers, ensuring that the maximum GPU temperature constraints will not violated based on the predicted load for that endpoint.

\myparagraph{Balancing IaaS and SaaS}
To enable SaaS workloads to balance airflow and power to reduce peaks, it is important to achieve a good balance of IaaS and SaaS workloads within each aisle and row.
We aim to place new VMs in an aisle and row that will not result in an excessive concentration of either IaaS or SaaS VMs.

\myparagraph{Migration}
Beyond initial VM placement, we can recalculate better placements and migrate VMs to address mispredictions or changes in workload behavior.
For SaaS workloads, we can create a new VM, transfer the workload, and then decommission the old VM.
However, for IaaS VMs, migration must be seamless and non-disruptive~\cite{clark2005live}.
Currently, live migration of GPU VMs is unsupported due to the complexities of GPU memory management, but this capability would enhance performance if implemented.

\subsection{Request Routing}
\label{sec:design-routing}

Once we place workloads in servers, \system{} leverages multi-instance SaaS endpoints to further smooth temperature and power draw through finely routing LLM inference requests.
We consider three constraint levels: aisle, row, and server.

\myparagraph{Aisle}
For each aisle, \system{} estimates the load on each server and calculates the total airflow demand for the VMs in that aisle.
This information is cached and recalculated every 5 minutes, updating whenever discrepancies are detected (\eg{}, if a server’s power consumption is higher than estimated).
\system{} then prevents routing requests to VMs that could trigger an airflow violation.

\myparagraph{Row}
Similarly to aisles, \system{} estimates the load for all servers in a row and aggregates these into a total power value.
It then assesses whether routing requests to VMs in that row would risk exceeding the peak power limit.

\myparagraph{Server}
\system{} tracks the current load and, using \Cref{eq:t_gpu}, estimates whether the GPU temperature for each server will exceed the threshold.
It avoids routing requests to VMs on servers with a high risk of overheating.

\subsection{Instance configuration}
\label{sec:design-config}

To ensure servers remain within cooling and power limits during load spikes or emergency events, \system{} calculates the maximum allowable airflow, GPU temperature, and server power for each instance.
Based on these limits, the \emph{Instance Configurator} uses the thermal and power profiles of the LLM from \Cref{fig:temp_power_perf_3d} to determine the optimal GPU frequency, batch size, model parallelism, quantization, and model size that maximize goodput without quality impact.

We also account for the time required to reassign these settings and prevent requests from being sent to instances during transitions~\cite{stojkovic2024dynamollm}.
For example, changing the model parallelism, size, or quantization level requires reloading the model, which can take a few seconds.
Given these overheads and the goal of maximizing quality, adjustments to model quantization and size are typically the last resort.

\subsection{Oversubscription and failures}

\myparagraph{Oversubscription}
Using \system{}, we can reduce the cooling and power requirements needed to run the same workload.
When the cloud provider initially provisions cooling and power for the datacenter, it typically plans for peak baseline demand.
As demand increases, the cloud operator can add more racks to the existing rows, utilizing the slack created by \system{}.
Additionally, our \system{} simulator can be used with an estimated workload to assess cooling and power requirements, enabling more precise provisioning.

\myparagraph{Failure management}
If there is a cooling or power failure, \system{} recalculates the new available airflow for each aisle, the power for each row, and the inlet temperature for each server.
Based on this, the request routing will steer requests away from constrained servers.
In addition, the \emph{Instance Configurator} will decrease the load accordingly.
If all these actions are not enough, \system{} applies regular power capping techniques to the IaaS VMs~\cite{polca}.

\subsection{Implementation}
\label{sec:design-implementation}

\myparagraph{Profiles}
\system{} includes an offline profiling phase that takes place during the initial stages of datacenter deployment, when operators run benchmarks and validation tests.
This phase models:
(1) the datacenter layout,
(2) the inlet temperature of each server,
(3) the temperature of each GPU,
(4) the server fan airflow, and
(5) the power-load for the servers.
In addition, when the provider onboards a new LLM, \system{} profiles the impact of each configuration parameter in that hardware, following the process described in \Cref{sec:characterization}.
During regular datacenter operation, \system{} refines these models on a weekly basis.
During this weekly update, \system{} collects power and load patterns for each server and row to predict their utilization.
We use a template-based approach that leverages data from the previous week~\cite{smartoclock}. We make these models and profiles available to the other three main components through regular data updates.

\myparagraph{VM Allocator}
We implement our workload placement policies in a rule-based \emph{VM Allocator}, inspired by Protean~\cite{hadary2020protean}, using three main rules:
(1) a validator rule filters aisles and rows based on peak airflow and power;
(2) a preference rule directs IaaS workloads to cooler servers and SaaS workloads to warmer servers.
For this rule, we categorize servers into three equally sized groups: cold, medium, and warm; and
(3) another preference rule guides VM placement based on the IaaS/SaaS balance, grouping servers into three categories: IaaS-heavy, SaaS-heavy, and balanced.
These rules use current cluster data and the weekly updated models and profiles.
Finally, our \emph{VM Allocator} applies these rules to select the final server to host the GPU VM.

\myparagraph{Load Balancer}
We deploy the SaaS endpoints on a dedicated set of VMs that expose an HTTP REST interface.
These VMs implement the \emph{Load Balancer}, which forwards LLM inference requests to the appropriate VM within the endpoint.
For each VM, \system{} evaluates the current VM state along with the server's thermal and power profiles to calculate the probability of violating any of the three operational limits (\ie{}, thermal, power, and performance).
Requests are not routed to VMs with a high risk of violation.

After the filtering step, \system{} applies state-of-the-art load balancing policies in the following order:
(1) route requests to instances that have previously handled requests from the same customer to maximize KV cache reuse~\cite{wu2024loongserve,preble2024efficient,gao2024cost};
(2) concentrate load to reduce energy consumption~\cite{stojkovic2024dynamollm}; and
(3) distribute requests across VMs to optimize performance.

\myparagraph{Instance Configurator}
To run the local LLM instances,
we use vLLM~\cite{vllm}, a state-of-the-art LLM inference platform. Note that 
\system{} can also integrate other platforms (\eg{}, TensorRT-LLM~\cite{tensorRTLLM}) with only minor interface modifications.
The LLM inference engine provides an HTTP REST API that receives requests from the \emph{Load Balancer}.

The local \system{} controller receives the updated thermal and power profiles for that server on a weekly basis.
This controller runs for every LLM iteration to estimate the optimal operational settings (including GPU frequency, batch size, model parallelism, model size, and quantization).
These computations are lightweight and cached for efficiency.
If needed, the controller updates the settings for each instance running on the VM.
It also restarts the LLM instance if changes to model parallelism, model size, or quantization are necessary.

\section{Evaluation}

\subsection{Methodology}

\myparagraph{Policies}
As a \emph{Baseline}, we use a thermal- and power-oblivious system with traditional VM placement~\cite{hadary2020protean} and LLM request routing~\cite{sarathi-serve} without any reconfiguration.
We implement \system{} as detailed in \Cref{sec:design-implementation} and evaluate six additional variations to assess the impact of each component, as well as their combinations, on VM placement (\emph{Place}), request routing (\emph{Route}), and instance configuration (\emph{Config}).

\myparagraph{Workload}
For VM arrivals, we use a one-week production trace from one of the A100 datacenters~\cite{a100} with a 50/50 split between IaaS and SaaS workloads.
This covers around one thousand servers with thousands of GPUs.
For SaaS LLM inference, we use Llama2~\cite{llama2} with the profile from \Cref{fig:temp_power_perf_3d}.
The requests are a subset from 10 endpoints, each with a VM count between 23 and 100 (\Cref{fig:char_servicedist}).

\myparagraph{Real cluster}
We conduct real experiments using a scaled-down version of the production trace, emulating two rows of 80 servers with a 50/50 IaaS/SaaS mix over one hour.
For SaaS, we run LLM inference across endpoints on Llama2~\cite{llama2}.
For IaaS, we use historical power readings directly.

\myparagraph{Simulation}
To evaluate \system{} at scale and compare policies under consistent conditions, we built a discrete-time simulator that models our datacenters as described in \Cref{sec:characterization-infra}.
This simulator replicates the load of IaaS VMs and the execution of LLM inference requests in SaaS VMs.

For cooling modeling, we use \Cref{eq:t_inlet,eq:t_gpu,eq:cfm}.
We evaluated various regression models, including random forests, multi-layer perceptrons, linear, polynomial, and piecewise polynomial regressions.
Piecewise polynomial achieved an MAE of <1\textdegree{}C, offering fast computation, efficient storage, and effective generalization for unseen values (\eg, random forests tend to overfit and struggle to predict temperatures lower than those in the training set).
These models simulate server temperatures based on IaaS power data and LLM inference requests for SaaS.

For power, the simulator uses real IaaS power readings and maps inference load to power consumption for each SaaS VM (\Cref{eq:power}).
It also tracks capping events by simulating their impact on both cooling and power infrastructure.

\subsection{\system{} operation}

\begin{figure}[t]
    \centering
    \includegraphics[width=\columnwidth]{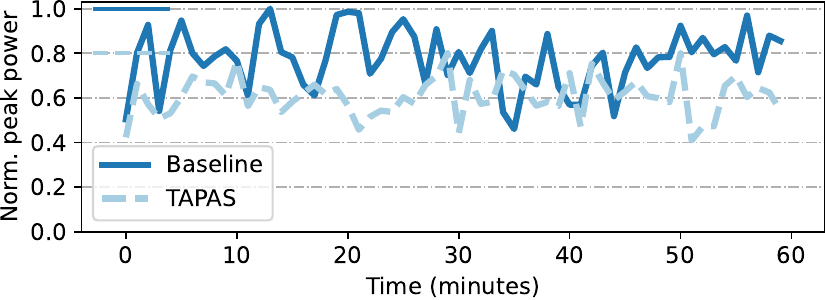}
    \vspace{-6mm}
    \caption{Peak power over 1-hour for \emph{Baseline} and \system{} while running real cluster experiments.}
    \label{fig:eval_emulation}
\end{figure}

\myparagraph{Real cluster}
\Cref{fig:eval_emulation} shows the peak row power for the 80 servers in the two rows measured at 1-minute intervals comparing the \emph{Baseline} and \system{}.
During regular operations, \system{} effectively reduces peak power, maintaining latency SLOs and result quality, achieving a 20\% reduction in peak utilization compared to \emph{Baseline}.
This experiment shows a 4\% absolute error compared to the simulation, validating the accuracy of our simulator.

\begin{figure}[t]
    \centering
        \includegraphics[width=\columnwidth]{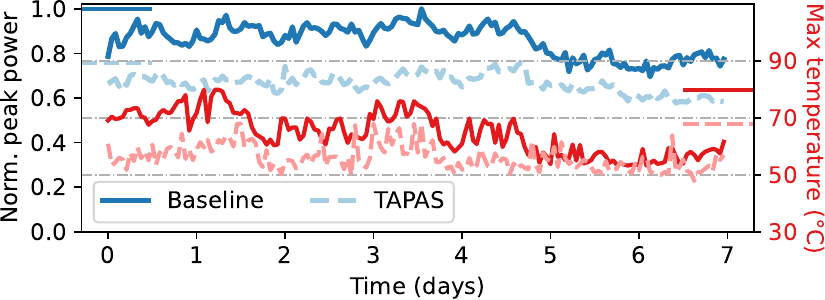}
        \label{fig:eval_week_simulation}
    \vspace{-6mm}
    \caption{Peak power and maximum temperature over 1-week for \emph{Baseline} and \system{} for large-scale simulations.}
    \label{fig:eval_week_power}
\end{figure}

\myparagraph{Simulation}
Extending to large-scale simulations, \Cref{fig:eval_week_power} shows the maximum temperature and row power over 5-minute intervals for one week.
Compared to \emph{Baseline}, \system{} reduces the maximum temperature by 15\% and peak power by 24\%, all without hurting result quality.

\begin{figure*}[t]
    \centering
    \includegraphics[width=\linewidth]{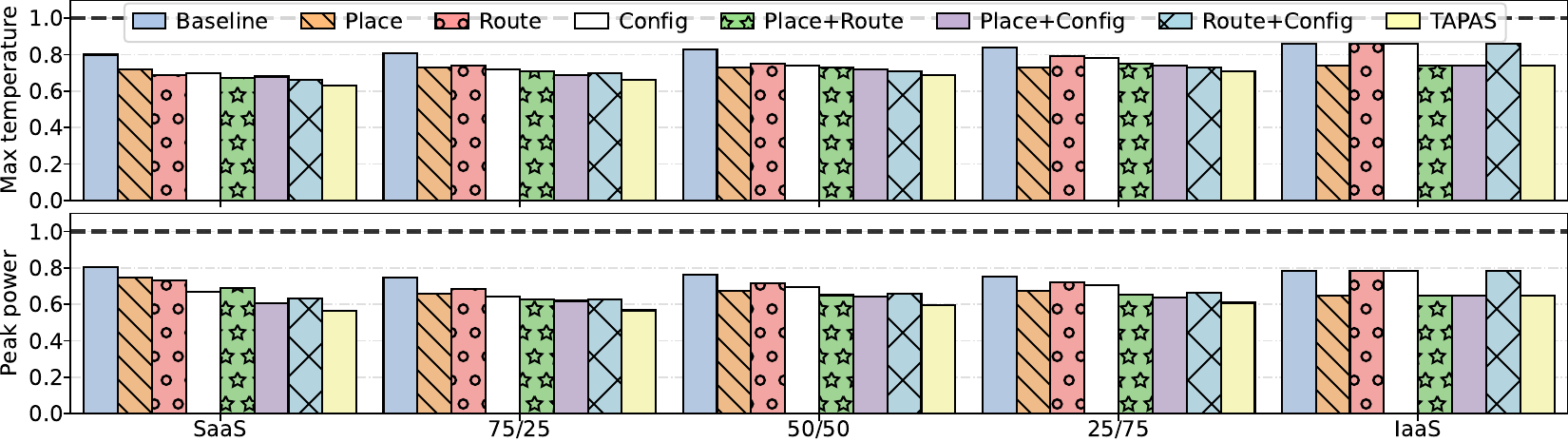}
    \caption{Normalized maximum temperature and peak power varying the policy and fraction of SaaS and IaaS workloads.}
    \label{fig:eval_savings}
\end{figure*}

\myparagraph{Ablation study}
\Cref{fig:eval_savings} shows the maximum temperature (top) and peak power (bottom) for the \emph{Baseline} and variations of \system{} over one week, normalized to the maximum provisioned values (indicated by the black lines).
Importantly, all policies operate without affecting quality or causing SLO violations under normal conditions.
For a 50/50 mix of IaaS/SaaS workloads (middle), each individual policy reduces both temperature and power by up to 12\% compared to the \emph{Baseline}, achieving these reductions by balancing or lowering the local load.
\emph{Place} performs slightly better, as it balances both IaaS and SaaS workloads across rows, while \emph{Route} and \emph{Config} focus on optimizing only SaaS workloads.
Although combining two components yields additional improvements, \system{} achieves the largest reductions in temperature and power (17\% and 23\%, respectively) through a holistic approach that integrates placement, routing, and configuration.

\myparagraph{Sensitivity to IaaS/SaaS fraction}
\Cref{fig:eval_savings} shows the maximum temperature and power varying the SaaS/IaaS fractions.
As expected, when the workload is entirely IaaS, \system{}’s effectiveness is limited to VM placement.
Conversely, \system{} achieves maximum reductions in temperature (23\%) and power (28\%) compared to the \emph{Baseline} when the workload is entirely SaaS, due to its flexibility.

\subsection{Oversubscription}
\begin{figure}[t!]
    \centering
    \includegraphics[width=\columnwidth]{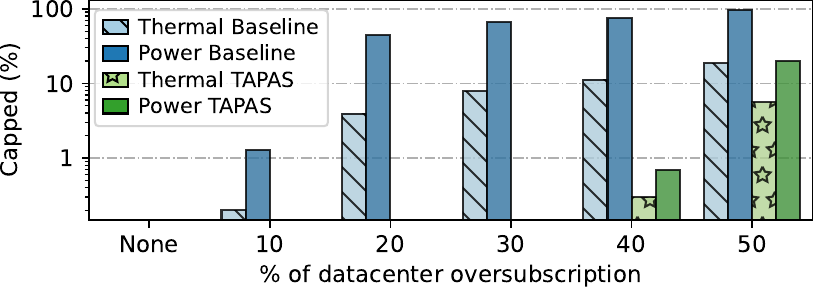}
    \caption{Time spent under thermal and power capping varying the oversubscription ratio for \emph{Baseline} and \system{}.}
    \label{fig:eval_oversub_temp_power}
\end{figure}

We analyze the effectiveness of \system{} in scenarios where the thermal and power infrastructures are oversubscribed.
\Cref{fig:eval_oversub_temp_power} shows the fraction of time during which thermal and power capping occurs as racks are added to the datacenter.
As expected, a datacenter without oversubscription (\emph{None}) experiences no capping due to thermal or power constraints under \emph{Baseline} and \system{}.
However, as additional servers are added, the \emph{Baseline} quickly begins to experience capping events, especially once oversubscription exceeds 20\%.
In contrast, \system{} supports up to 40\% more servers without impacts on quality of results while maintaining thermal and power capping below 0.7\% of the time, enabling safe thermal and power oversubscription.

\subsection{Failure management}
In the event of thermal (AHU) or power (UPS) failures, datacenters must immediately adapt to reduced capacity limits of 90\% and 75\%, respectively.
\Cref{tab:emergency_comparison} compares the performance and quality impact on \emph{Baseline} and \system{} over a 5-minute peak load period.
As precise IaaS performance impact is challenging to measure, we present the effects on both IaaS and SaaS workloads as the percentage of frequency capped relative to maximum frequency, adjusted by the fraction of workloads affected.
To stay within constraints, \emph{Baseline} applies uniform frequency caps up to 35\% across servers, leading to significant performance drops.
In contrast, \system{} maintains (or even improves) performance with up to 12\% quality impact (\ie, accuracy drop by number of requests directed to smaller models).
\system{} effectively manages temperature and power through selective actions, such as routing requests to smaller models only when necessary.

\begin{table}[t!]
    \centering
    \footnotesize
    \begin{tabular}{c cc|cc||cc|cc}
        \toprule
        & \multicolumn{4}{c||}{Power Emergency} & \multicolumn{4}{c}{Thermal Emergency} \\
        \cline{2-9}
        & \multicolumn{2}{c|}{Baseline} & \multicolumn{2}{c||}{\system{}} & \multicolumn{2}{c|}{Baseline} & \multicolumn{2}{c}{\system{}} \\
        \cline{2-9}
        & IaaS & SaaS & IaaS & SaaS & IaaS & SaaS & IaaS & SaaS \\
        \midrule
        Perf & \textcolor{red}{-35\%} & \textcolor{red}{-28\%} & 0\% & +16\% & \textcolor{red}{-22\%} & \textcolor{red}{-19\%} & 0\% & +10\% \\
        Quality & 0\% & 0\% & 0\% & \textcolor{red}{-12\%} & 0\% & 0\% & 0\% & \textcolor{red}{-6\%} \\
        \bottomrule
    \end{tabular}
    \caption{Comparison of \emph{Baseline} and \system{} in power and thermal emergencies across IaaS and SaaS}
    \vspace{-3mm}
    \label{tab:emergency_comparison}
\end{table}
\section{Related work}

\myparagraph{Datacenter cooling management}
Researchers~\cite{jiang2019isca, coolEdge, cooldc, coolair, coolprovision, geng2024tesla} proposed adaptive cooling systems to address thermal hotspots and enhance cooling efficiency through optimized thermal control with various technologies (\eg{}, warm water~\cite{coolEdge}, immersion-cooling~\cite{cooldc}, and free-cooling~\cite{coolair}).
CoolProvision~\cite{coolprovision} optimizes under-provisioned cooling while maintaining performance.
Instead, \system{} reduces hotspots in LLM inference clusters through VM placement, request routing, and instance configuration.

\myparagraph{Thermal-aware scheduling}
Prior work optimizes datacenter job placement to reduce thermal issues~\cite{rt-tas, vmt, thermal-aware-scheduling, CHEN2023578, vasic2010icac, tp-hadoop}.
For example, RT-TAS~\cite{rt-tas} proposes a thermally-balanced task-to-core assignment for integrated GPU-CPU platforms while PTDS~\cite{CHEN2023578} optimizes VM-to-host scheduling to prevent hotspots and reduce cooling energy.
However, traditional thermal- or power-aware scheduling approaches yield suboptimal results in LLM serving due to its unique challenges, as discussed in \Cref{sec:characterization}.

\myparagraph{Datacenter power management}
To improve datacenter power utilization, Flex~\cite{flexDatacenter} safely oversubscribes reserved power through offline workload placement and online load shedding.
SmartOClock~\cite{smartoclock} distributes power budgets efficiently through power predictions for workload-aware overclocking.
SmoothOperator~\cite{smoothOperator} spreads services with synchronous power patterns evenly across the datacenter to reduce peak power draw.
\system{} contributes to power utilization improvement focusing on LLM inference clusters.

\myparagraph{LLM serving}
Recent works explore unique GPU and LLM serving characteristics~\cite{vspetko2021dgx, towardsGreen, wordstowatts}, challenges~\cite{li2024unseen}, and opportunities~\cite{stojkovic2024dynamollm} for power and energy.
POLCA~\cite{polca} introduces a power oversubscription framework for LLM inference clusters.
$\mu$-Serve~\cite{qiu2024muserve} enables power-aware LLM serving through model parallelism and predictive request scheduling.
Other orthogonal LLM serving optimizations to \system{} include key-value cache management~\cite{vllm}, continuous batching~\cite{orca}, scheduling~\cite{aiops2024qiu, wu2023fast}, autoscaling~\cite{kakolyris2024slo}, prefill-decode interference reduction~\cite{splitwise, zhong2024distserve, sarathi-serve}, hardware heterogeneity~\cite{ye2024deep, weng2022mlaas,wilkins2024hybrid,splitwise}, and geographical load balancing~\cite{li2024towards}.

\section{Conclusions}
We introduced \system{}, a system for thermal- and power-aware scheduling of LLM inference in GPU datacenters, leveraging VM placement, request routing, and instance configuration, while maintaining performance and quality.
\system{} maximizes cooling and power efficiency with minimal quality impact, effectively reducing thermal/power peaks, supporting oversubscription, and handling failures.

\bibliographystyle{plain}
\bibliography{references}

\end{document}